\documentstyle[12pt,chicago]{article}
\input epsf

\newcommand{\cov}{{\rm Cov}}
\newcommand{\var}{{\rm Var}}
\newcommand{\ex}{{\rm E}}

\newcommand{\trace}{{\rm Tr}}

\newcommand{\trans}{^{T}}

\newcommand{\ci}{\amalg}

\newcommand{\be}{\begin{eqnarray}}
\newcommand{\ee}{\end{eqnarray}}
\newcommand{\beq}{\begin{equation}}
\newcommand{\eeq}{\end{equation}}
\newcommand{\bd}{{\rm [B/D] }}

\title{Bayes linear adjustment for variance matrices}
\author{
Darren J Wilkinson
\thanks{E-mail: {\tt d.j.wilkinson@durham.ac.uk} --- WWW: {\tt
    http://fourier.dur.ac.uk:8000/djw.html} --- \bd information can be
  obtained from {\tt http://fourier.dur.ac.uk:8000/stats/bd/}} \\
\and
Michael Goldstein \\
}
\date{Department of Mathematical Sciences,  University of Durham,\\
Science Laboratories, South Road, \\
Durham  DH1 3LE, England.}

\begin{document}
\maketitle
\setcounter{footnote}{2}

\begin{abstract}
We examine the problem of covariance belief revision
using a geometric approach. We 
exhibit an inner-product space where covariance matrices live naturally ---
a space of random real symmetric matrices.
The inner-product on this space captures aspects of our beliefs
about the relationship between covariance matrices of interest to us,
providing a structure rich enough for us to adjust
beliefs about unknown matrices in the light of data
such as sample covariance matrices, exploiting second-order
exchangeability specifications.
\end{abstract}

\noindent{\it Keywords:\/} BELIEF ADJUSTMENT; COVARIANCE ESTIMATION;
EXCHANGEABILITY; LINEAR BAYES; MATRIX INNER-PRODUCT; SUBJECTIVIST.

\section{Revising beliefs about covariance structures}

Quantifying
relationships between variables is of fundamental importance in
Bayesian analysis. However, there are
many difficulties associated even with learning about covariances. 
For example, it
 is often difficult to make prior covariance specifications, but
 it is usually even harder to make the statements about the uncertainty in
  these covariance statements which are required in order to learn about the
  covariance statements from data.
 Further, a covariance 
structure is more than just a collection of random
  quantities, so we should aim to analyse such structures in a space
  where they live naturally. In this paper, we develop and illustrate
  such an approach, based around a geometric representation for
  variance matrices and exploiting second-order exchangeability 
specifications for them.

\section{Current approaches to covariance estimation}

Until recently, most authors have followed a Wishart conjugate prior
approach (see for example, \citeN{ccdispmat} or \citeN{lhcovmat}).
This approach, whilst tractable, places severe
restrictions on the form of the prior distribution.
More recently, a different approach has been proposed by 
\citeN{tlcovmat}, who learn about the 
log of the covariance
matrix using data. This solves the
positivity problems associated with covariance revision,
but makes prior belief specification more difficult.

\citeN{blzicm}, make further progress: working within a distributional
Bayesian paradigm, they develop a reasonably flexible prior
over the elements of
a covariance structure,  and offer interpretations
for the parameters that one 
is required to specify. However, this work is still 
restricted to multivariate Normal likelihoods, 
and there is a weak
restriction on the form of the mean structure for the data.

\section{Bayes linear methods}

 The Bayes linear approach to subjective statistical
  inference makes expectation (rather than probability)
  primitive.
 An overview of the methodology is given in \citeN{fgcross}.
  In particular, as we are not forced to specify full prior
  measures over all variables of interest, we may exploit second-order 
exchangeability to allow us to construct statistical models directly
from small numbers of belief specifications 
  over observables. Foundational issues raised by Bayes linear
analysis of exchangeable specifications are
  discussed in \citeN{mgrevexch}. We now show that
these methods offer a simple and tractable approach to covariance 
estimation,
linking sample covariance matrices with their
  ``population'' counterparts, in a natural geometric setting.

\section{Exchangeable representations for covariances}

Let $\underline{X}_1,\underline{X}_2,\ldots$ be an infinite,
second-order exchangeable sequence of random vectors, each of length
$r$, namely a sequence for which $\underline{X}_k=(X_{1k},\ldots,X_{rk})\trans$,
$\mu_i=\ex(X_{ik}), c_{ij}=\cov(X_{ik},X_{jk})$ does not depend on
$k$, and $c_{ij}^\prime=\cov(X_{ik},X_{jl}), k\not= l$ does not
depend on $k, l$.

 From this specification, we may use the second-order exchangeability
 representation 
theorem \cite{mgexchbel} to decompose $X_{ik}$ as
\beq
X_{ik} = M_i + R_{ik} \label{exrep}
\eeq
where $ \ex(R_{ik}) = \cov(M_j,R_{ik}) = \cov(R_{ik},R_{jl})=0,
\forall i,j, k\not=l$, and
the vectors $\underline{R}_{k}=(R_{1k},\ldots,R_{rk})\trans$
 form a second order exchangeable sequence.
Here, $M_i$ may be thought of as representing underlying population
behaviour, and $R_{ik}$ as representing individual variation.

Consider the 
sequence of $\frac{r(r+1)}{2}$-dimensional
vectors 
\beq
\underline{Y}_k =
(R_{1k}R_{1k},\ldots,R_{1k}R_{rk},R_{2k}R_{2k},\ldots,R_{2k}R_{rk},\ldots,\ldots,R_{rk}R_{rk})\trans
\eeq
representing
 the quadratic products of
  the residuals. Suppose that we assume that the $\underline{Y}_k$ are
  second-order exchangeable, and that we express the additional specifications
$v_{ijpq}=\cov(R_{ik}R_{jk},R_{pk}R_{qk})$ and  $v_{ijpq}^\prime=$ 
$\cov(R_{ik}R_{jk},R_{pl}R_{ql}), k\not=l$. Then we
  may similarly decompose the elements of $\underline{Y}_k$ as
\beq
R_{ik}R_{jk} = V_{ij} + U_{ijk} \label{quadrep}
\eeq
with properties as for representation (\ref{exrep}). In particular
$\cov(U_{ijk},U_{pqk})=u_{ijpq}$ is not dependent on $k$. Here, 
$V_{ij}$ represents underlying covariance behaviour, and $U_{ijk}$
represents individual variation within the quadratic products of residuals.

If we observe a sample $\underline{X}_1,\ldots,\underline{X}_n$ of
size $n$, then
sample covariances take the form
\be
S_{ij} &=& \frac{1}{n-1}\sum_{w=1}^n(X_{iw}-X_{i\cdot})(X_{jw}-X_{j\cdot}) \\
&=& \frac{1}{n-1}\sum_{w=1}^n (R_{iw}-R_{i\cdot})(R_{jw}-R_{j\cdot}) \label{sres}
\ee 
 Beliefs over the sample covariances $S_{ij}$ are, by (\ref{sres}),
 uniquely determined by  representation (\ref{quadrep}), and can be written
\beq
S_{ij} = V_{ij} + T_{ij}
\eeq
where $V_{ij}$ is as in (\ref{quadrep}), $\ex(T_{ij})=0$ and
$\cov(V_{ij},T_{ij})=0$.
The covariance structure over $T_{ij}$ is given by
\beq
\cov(T_{ij},T_{pq}) =\frac{u_{ijpq}}{n}.
\eeq
Observing sample covariances from a sample of size $n$ reduces 
uncertainty for $V_{ij}$, the underlying
covariance values, but is uninformative for the $U_{ijk}$ for $k>n$.

Let $S$ be the matrix whose $(i,j)$th element is $S_{ij}$, and define
$V$ and $T$ similarly.
We then have
\beq
S=V+T
\eeq

\section{Geometric representation for random matrices}

We now develop the representation which will allow us to treat a
covariance matrix as a single object.
Let 
\(
B=[B_1,B_2,\ldots]
\)
 be a collection of random $r\times r$ real symmetric matrices, 
representing unknown matrices of interest to us. These might, for
example, represent
population covariance matrices. Let
\(
D=[D_1,D_2,\ldots]
\)
be another such collection, representing observable
 matrices (such as
sample covariance matrices). Finally, let
\(
C=[C_1,C_2,\ldots]
\)
be a basis for the space of constant $r\times r$ real symmetric 
matrices. We now form a vector space
\beq
L=span\{ B\cup C \cup D \}
\eeq
of all linear combinations of the elements of these collections,
and define the inner-product (over equivalence classes) on $L$ as
\beq
 (P,Q) = \ex(\trace(PQ)) \quad  \forall P,Q\in L
\label{ip}
\eeq
which induces the metric
\beq
d(P,Q)^2 =
  \ex( \Vert P-Q \Vert_F^2 ) \quad \forall P,Q\in L,
\eeq
where $\Vert\cdot\Vert_F$ denotes the Frobenius norm of a matrix. This 
is the sum of the squares of the elements, or equivalently, the
  sum of the squares of the eigenvalues.
Where necessary, we form the completion of the space. 
The (complete) inner-product space is 
denoted by $M$.

Analogously with the revision of belief over scalar quantities \cite{mgrevprev},
we 
learn about the elements of the collection $B$, by orthogonal 
projection into subspaces of $M$ spanned by
elements of the collection $C\cup D$, in order 
to obtain the corresponding adjusted expectations, 
namely the linear combinations of sample covariance matrices which 
give our adjusted beliefs.

If all matrices of interest contain only one non-zero component 
(all in the same position), the inner product becomes
$(P,Q)=\ex(P_{ij}Q_{ij})$, inducing the distance
$d(P,Q)^2=\ex((P_{ij}-Q_{ij})^2)$, as for the usual Bayes linear
theory for scalar quantities. The matrix structure is a
generalisation of the scalar Bayes linear structure, and scalar Bayes
linear adjustments can be recovered by decomposing all variance
structures to the one component level.

The matrices we are considering do not have to be finite
dimensional. All of the theory remains valid if we think in terms of
representations of random linear self-adjoint operators on a
(possibly infinite-dimensional) vector space.

\section{Decomposing the variance structure}

As a simple example, $B$ might consist only of the 
``population" covariance matrix, $V$ for a particular problem, and
$D$ might be the corresponding sample covariance matrix, $S$, based on
$n$ observations. 
In this case, our adjusted
  expectation for the ``population'' matrix would be a weighted
  linear combination of the prior and sample covariance matrices. However,
 by breaking down the sample covariance
  matrix into its component sub-matrices, we may
  resolve a greater proportion of our uncertainty about 
the ``population'' covariance
  matrix.

For simplicity, consider the case where we wish to learn about the 
covariance structure induced by representation (\ref{quadrep}) 
for $2$-dimensional vectors. The covariance matrices will be $2\times
2$. Consider the sample
covariance matrix
\beq
S=
\left(
\begin{array}{cc}
S_{11} & S_{12} \\
S_{12} & S_{22}
\end{array}
\right)
\eeq
and the corresponding ``population" covariance matrix
\beq
V=
\left(
\begin{array}{cc}
V_{11} & V_{12} \\
V_{12} & V_{22}
\end{array}
\right)
\eeq
In the notation of the previous section, we could restrict ourselves to
\beq
B=[V], D_S=[S], C=\left[ 
\left(
\begin{array}{cc}
1 & 0 \\
0 & 0 
\end{array}
\right)
,
\left(
\begin{array}{cc}
0 & 1 \\
1 & 0 
\end{array}
\right)
,
\left(
\begin{array}{cc}
0 & 0 \\
0 & 1 
\end{array}
\right)
 \right],
\eeq
where all $2\times 2$ matrices can be constructed as linear
combinations of the elements of $C$. Using these collections, 
our adjusted expectation for $V$ given $D_S$ would take the form
\beq
\ex_{D_S}(V) = (1-\alpha )\ex(V) + \alpha S
\eeq
where $\alpha$ is the coefficient of the orthogonal projection
 determined by the inner-product (\ref{ip}). Explicitly:
\be
\alpha &=& \frac{(V-\ex(V),V-\ex(V))}{(S-\ex(S),S-\ex(S))} \\
&=& \frac{ \sum_{i=1}^2\sum_{j=1}^2 n\var(V_{ij})}{
  \sum_{i=1}^2\sum_{j=1}^2 \{ n\var(V_{ij})+\var(U_{ij}) \} }
\ee

However, to improve the precision of our estimates, our projection 
space could be enlarged by constructing
\beq
D_I=\left[
\left(
\begin{array}{cc}
S_{11} & 0 \\
0 & 0 
\end{array}
\right)
,
\left(
\begin{array}{cc}
0 & S_{12} \\
S_{12} & 0 
\end{array}
\right)
,
\left(
\begin{array}{cc}
0 & 0 \\
0 & S_{22} 
\end{array}
\right)
\right]
\eeq
We call such a space the {\it individual variance collection\/}. This
allows different sample covariances to have different weights, if for
example, we have higher prior uncertainty about some of the variances. 
Indeed, we may take this a stage further, and construct
\be
D_C&=&\left[
\left(
\begin{array}{cc}
S_{11} & 0 \\
0 & 0 
\end{array}
\right)
,
\left(
\begin{array}{cc}
0 & S_{11} \\
S_{11} & 0 
\end{array}
\right)
,
\left(
\begin{array}{cc}
0 & 0 \\
0 & S_{11} 
\end{array}
\right),
\right.
\nonumber\\
&&
\left(
\begin{array}{cc}
S_{12} & 0 \\
0 & 0 
\end{array}
\right)
,
\left(
\begin{array}{cc}
0 & S_{12} \\
S_{12} & 0 
\end{array}
\right)
,
\left(
\begin{array}{cc}
0 & 0 \\
0 & S_{12} 
\end{array}
\right),
\nonumber\\
&&
\left.
\left(
\begin{array}{cc}
S_{22} & 0 \\
0 & 0 
\end{array}
\right)
,
\left(
\begin{array}{cc}
0 & S_{22} \\
S_{22} & 0 
\end{array}
\right)
,
\left(
\begin{array}{cc}
0 & 0 \\
0 & S_{22} 
\end{array}
\right)
\right]
\ee
We call this last collection the {\it complete variance
  collection\/}.
This not only allows the different covariances to have different 
weights, but also allows relationships between covariances to
have an effect on the adjustment.
If we project $V$ into $D_C$, then our adjusted expectation for $V$
will correspond precisely with the 
adjustment which would have been obtained using Bayes linear 
estimation on the quadratic products of the residuals in the
scalar space.

We can break down the population matrix in the same way if necessary.
In particular, we let
\beq
V_I=\left[
\left(
\begin{array}{cc}
V_{11} & 0 \\
0 & 0 
\end{array}
\right)
,
\left(
\begin{array}{cc}
0 & V_{12} \\
V_{12} & 0 
\end{array}
\right)
,
\left(
\begin{array}{cc}
0 & 0 \\
0 & V_{22} 
\end{array}
\right)
\right].
\eeq

As we enlarge the projection space, we resolve more of our uncertainty
about the variance structures, at the expense of doing more work.
Generally we should project into as rich
a space as is practicable, but for large variance matrices, the difference
both in computational effort and in effort required for prior specification,
between adjusting by $D_S$, $D_I$ and $D_C$ is substantial, so that we
must make a subjective assessment of the relative benefits
of each adjustment.

\section{Example}
\subsection{Examination performance}

 We are currently investigating the examination performance of 
 first year mathematics undergraduate
students at Durham university. We are particularly interested in those
  students who have only one A Level in mathematics, and so we
  restrict attention to these in our account.
 For illustrative purposes, we focus on a few key variables, namely
  a summary of A Level performance ($A$), performance in
  the Christmas exams ($C$), and the end of year exam average ($E$).

\begin{figure}
\epsfysize=2.5in
\hspace{2in}\epsfbox[0 0 270 270]{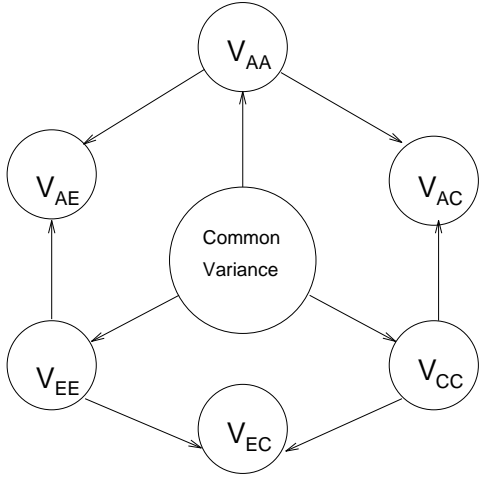}
\caption{A conditional linear independence graph for the 
mean components of the quadratic products of the residuals}
\label{hex}
\end{figure}

For the exchangeable decomposition
of (say) $A_k$, we will write
\beq
A_k = M_A + R_{A_k}
\eeq
and for the exchangeable decomposition of (say)
$R_{A_k}R_{C_k}$, we write
\beq
R_{A_k}R_{C_k} = V_{AC} + U_{AC_k}
\eeq
so that, for example, $V_{AC}$ represents the underlying covariance
between the $A$ and $C$ variables, and $U_{AC_k}$ represents the
residual for the $k$th observation.
We construct the ``population'' and sample covariance matrices:
\beq
V = \left(\begin{array}{ccc}
V_{AA} & V_{AE} & V_{AC} \\
V_{AE} & V_{EE} & V_{EC} \\
V_{AC} & V_{EC} & V_{CC}
\end{array}\right) 
,\quad 
S = \left(\begin{array}{ccc}
S_{AA} & S_{AE} & S_{AC} \\
S_{AE} & S_{EE} & S_{EC} \\
S_{AC} & S_{EC} & S_{CC} \\
\end{array}\right)
\eeq
A conditional linear independence graph \cite{mginfl} was 
formed to represent
beliefs about the relationships between the quadratic products of
the residuals (Figure \ref{hex}). 
The common variance node reflects beliefs about the positive
correlation between variances. Covariances are influenced by the
corresponding variances. 
 This graph was used to help structure the 
belief specification over the mean components
of the variance structure.

Specifications are also required over the residual components of the
variance structure. These
specifications are more difficult to make, since
we are not used to thinking about such quantities. In this example,
for simplicity, our belief specifications over the residual structure
were chosen to be consistent with those imposed under a multivariate
normal specification corresponding to our prior specifications over
the elements $R_{ik}$.
Having made specifications over the quadratic products of residuals,
beliefs over all relevant covariance matrices are now determined.

From the sample covariance matrix, $S = D_S$, we construct the
individual variance collection, $D_I$ ($6$ objects)
and the complete variance collection, $D_C$ ($36$ objects), as 
well as the
individual collection for the mean structure, $V_I$ ($6$ objects).
 We form the random matrix space, $M$ over all these objects, and
investigate adjustments in this space.

\subsection{Quantitative analysis}

The prior covariance matrix was specified directly as follows:
\beq
\ex(V)=
\left(
\begin{array}{rrr}
   7.98 & 11.14 & 15.75 \\
   11.14 & 56.26 & 53.04 \\
   15.75 & 53.04 & 100.00 
\end{array}\right)
\eeq
The sample covariance matrix (34 cases) is:
\beq
S=
\left(\begin{array}{rrr}
    8.28 & 20.15 & 24.75 \\
   20.15 & 178.30 & 160.74 \\
   24.75 & 160.74 & 258.26
\end{array}\right)
\eeq
The adjusted matrices were formed as the appropriate linear combinations
of the observables, as described in section 5, and derived explicitly 
for the simplest case in section 6.
\be
\ex_{D_S}(V)&=&
\left(\begin{array}{rrr}
   8.08 & 14.08 & 18.69 \\
   14.08 & 96.08 & 88.18 \\
   18.69 & 88.18 & 151.65 
\end{array}\right) \\
\ex_{D_I}(V)&=&
\left(\begin{array}{rrr}
8.04 & 15.96 & 17.72 \\
15.96 & 98.90 & 78.63 \\
17.72 & 78.63 & 159.21
\end{array}\right) \\
\ex_{D_C}(V)&=&
\left(\begin{array}{rrr}
8.30 & 15.43 & 20.06 \\
15.43 & 92.04 & 80.66 \\
20.06 & 80.66 & 156.79
\end{array}\right)\label{ecv}
\ee
These
adjusted matrices may be used as a basis for assessing our 
posterior beliefs about the matrix object \cite{mgrevexch}. 

Note that the last matrix (\ref{ecv}) represents the 
adjusted 
matrix which would have been obtained using a standard Bayes linear 
analysis on the quadratic products of the residuals. In this 
particular example, all adjusted matrices are 
positive definite. In general, we view negative eigenvalues in the
revised structure as providing diagnostic warnings of possible
conflicts between prior beliefs and the data.

We would like to be able to compare the estimates of
$V$: $\ex_{D_C}(V)$, $\ex_{D_S}(V)$, and $\ex_{D_I}(V)$. 
Thus, we use the standard interpretive and diagnostic features of
the Bayes linear methodology
to assess the model and understand the adjustments taking place.

\subsection{Bayes linear influence diagram}

\begin{figure}[thbp]
\hspace{0.5in}\epsfbox{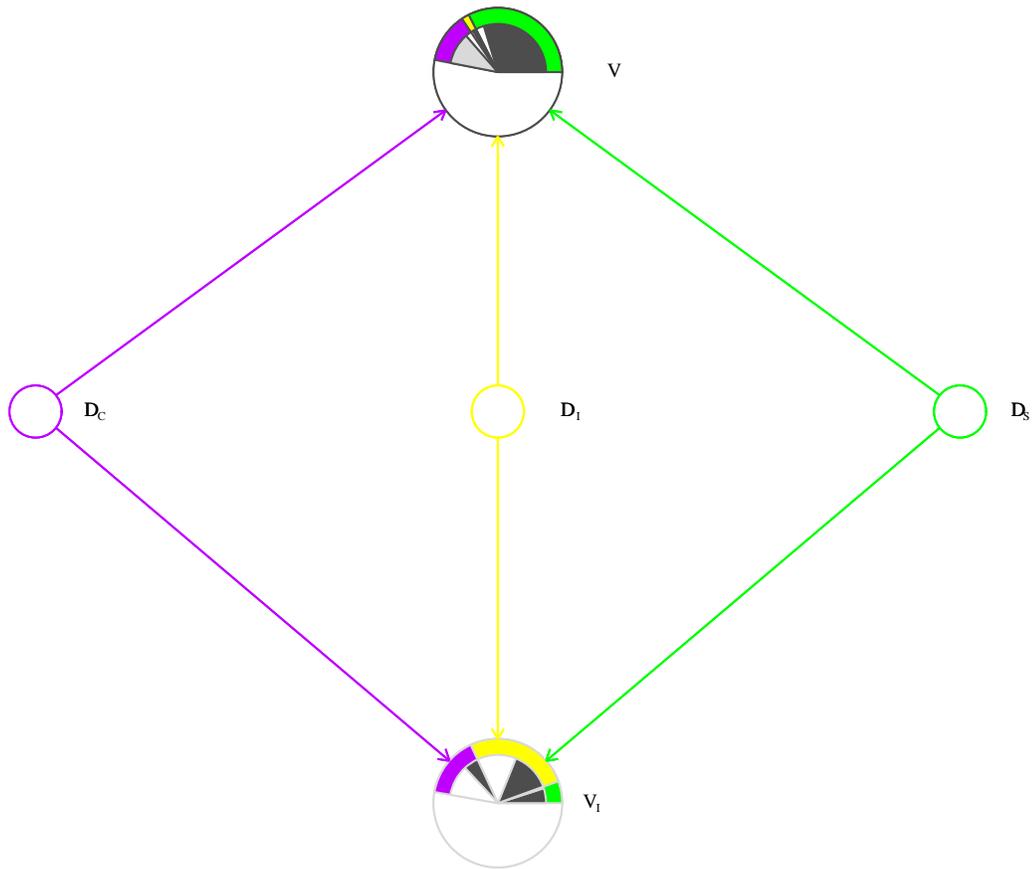}
\caption{Diagnostic influence diagram summarising changes in expectation
  of the matrix objects}
\label{mainid}
\end{figure}

Figure \ref{mainid} shows a Bayes linear influence diagram representing the
 adjustments and corresponding
diagnostic information for the random matrices.
 Such diagrams are 
described in detail in
\citeN{gfsbeer} for random quantities, with a similar
interpretation for random matrices, where conditional
linear independence is determined instead by the 
inner-product (\ref{ip}), so
that conditional linear independence becomes
\beq
B \ci C / D \iff \ex[\trace\{  ( B - \ex_D(B) )( C - \ex_D(C))  \}] = 0
\eeq
as described in \citeN{mginfl}.

 The outer shadings of the $V$ node represent proportions of
uncertainty about $V$ resolved by projection 
into the various spaces. Shadings
start at 3 o'clock, and progress in an anti-clockwise fashion. The
full circle represents the total uncertainty about the value of the
covariance matrix. The first outer portion shaded represents the
proportion of our uncertainty resolved by the sample covariance
matrix alone ($D_S$). By comparing this with the first shaded portion for
the $V_I$ node, we see that we have learned 
considerably more about the matrix object, than we have 
about the 6-dimensional space over the individual variance collection.

The next shading gives the {\it additional\/} information gained by using
the individual collection as the projection space. We see that this tells
us a great deal more about the elements of the $V_I$ collection, but
little about the matrix object as a whole.
The other shading shows the additional uncertainty 
resolved due to including the complete variance collection in our 
projection space. We see
that there is information to be gained by 
enriching our projection space, but 
we must balance information gained with extra effort involved.
Whether or not we choose to include the complete variance collection will
depend upon the size of the problem under consideration, and upon how
much the answer really matters.

Shadings in the centres of the nodes are diagnostics based
on the {\it size\/} and {\it bearing\/} of the adjustments, as described in
\citeN{mgtraj}. We generalise the bearing to the space
of random matrices as follows:
For any given constant matrix, $G$, and projection space $D$,
the bearing is defined
to be the unique random matrix, $B$, with
the property
\beq
(A-\ex(A),B) = (\ex_d(A),G) - (\ex(A),G) ,\ \forall A \in M
\eeq
where $\ex_d(A)$ represents the realisation of $\ex_D(A)$
after observing $D=d$.
Different choices of the constant
matrix, $G$, give information about different projections of the
adjusted expectations. 
The choice of $G$ which causes diagnostics to match up exactly with those
for scalar Bayes linear adjustment in the case where we are dealing
exclusively with one-component matrices, is the choice given by the constant
matrix whose elements are all $1$.
At the centre of the $V$ node, dark (light) 
shadings represent changes in expectation
  larger (smaller) than we expected {\it a priori\/}.
We can see that adjusting by the sample covariance matrix, $D_S$,
 caused a much larger 
change in expectation than we expected {\it a priori\/}. This is
evidence that we were too confident about our ability to predict
the true value of the covariance matrix, and suggests that
we should re-examine the prior specification.
We also notice that adding the complete variance collection, $D_C$,
 to the
adjustment had the potential to change our expectation considerably,
but in fact,  hardly changed it at all. This is perhaps evidence that we
overestimated the importance of the covariance terms.

\section{Summary}

Analysing matrices in a space where they live naturally not only has 
great aesthetic appeal, but is very powerful and illuminating in
practice.
Working in this space simplifies the handling of 
large matrices, by reducing the number of quantities involved
and summarising effects over the whole covariance structure.
For the same reasons, diagnostic information about
adjusted beliefs is easier to
 interpret. We may decompose structures as much or
as little as we wish.

This approach allows us to learn about collections of covariance
structures, and examine their relationships.
 It generalises the ``element by element'' approach to
  revision, which can be viewed
  as taking place in a subspace of the larger space.
Exchangeability representations lie at the heart of the methodology:
all of our specifications are over observables, or quantities
constructed from observables, rather than artificial 
  model parameters, and we make no distributional assumptions for the
data or the prior.

\section{Acknowledgements}

The first named author is supported by a grant from the UK's EPSRC.
All  computations, and the production of the diagnostic influence 
diagram, were carried out using the Bayes linear
 computing package, \bd, outlined in 
\citeN{bdworks}, and explained in detail in \citeN{gwblincomp}. The
comments from an anonymous referee have helped us improve
the clarity of this paper.

% Reference Page

\bibliographystyle{chicago}
\bibliography{../../bib/bayeslin,../../bib/djw}

\begin{thebibliography}{}

\bibitem[\protect\citeauthoryear{Brown, Le, and Zidek}{Brown
  et~al.}{1994}]{blzicm}
Brown, P., N.~Le, and J.~Zidek (1994).
\newblock Inference for a covariance matrix.
\newblock In P.~Freeman and A.~Smith (Eds.), {\em Aspects of Uncertainty: A
  Tribute to D.V.~Lindley}, pp.\  77--92. Wiley.

\bibitem[\protect\citeauthoryear{Chen}{Chen}{1979}]{ccdispmat}
Chen, C. (1979).
\newblock Bayesian inference for a normal dispersion matrix.
\newblock {\em J. Roy. Statist. Soc. Ser. B\/}~{\em 41}, 235--248.

\bibitem[\protect\citeauthoryear{Farrow and Goldstein}{Farrow and
  Goldstein}{1993}]{fgcross}
Farrow, M. and M.~Goldstein (1993).
\newblock {B}ayes linear methods for grouped multivariate repeated measurement
  studies with application to crossover trials.
\newblock {\em Biometrika\/}~{\em 80\/}(1), 39--59.

\bibitem[\protect\citeauthoryear{Goldstein}{Goldstein}{1981}]{mgrevprev}
Goldstein, M. (1981).
\newblock Revising previsions: a geometric interpretation.
\newblock {\em J. R. Statist. Soc.\/}~{\em B:43}, 105--130.

\bibitem[\protect\citeauthoryear{Goldstein}{Goldstein}{1986}]{mgexchbel}
Goldstein, M. (1986).
\newblock Exchangeable belief structures.
\newblock {\em J. Amer. Statist. Ass.\/}~{\em 81}, 971--976.

\bibitem[\protect\citeauthoryear{Goldstein}{Goldstein}{1988}]{mgtraj}
Goldstein, M. (1988).
\newblock The data trajectory.
\newblock In J.-M. Bernardo et~al. (Eds.), {\em {B}ayesian Statistics 3}, pp.\
  189--209. Oxford University Press.

\bibitem[\protect\citeauthoryear{Goldstein}{Goldstein}{1990}]{mginfl}
Goldstein, M. (1990).
\newblock Influence and belief adjustment.
\newblock In J.~Smith and R.~Oliver (Eds.), {\em Influence Diagrams, Belief
  Nets and Decision Analysis}. Chichester: Wiley.

\bibitem[\protect\citeauthoryear{Goldstein}{Goldstein}{1994}]{mgrevexch}
Goldstein, M. (1994).
\newblock Revising exchangeable beliefs: subjectivist foundations for the
  inductive argument.
\newblock In P.~Freeman and A.~Smith (Eds.), {\em Aspects of {U}ncertainty: A
  Tribute to D. V. Lindley}. Wiley.

\bibitem[\protect\citeauthoryear{Goldstein, Farrow, and Spiropoulos}{Goldstein
  et~al.}{1993}]{gfsbeer}
Goldstein, M., M.~Farrow, and T.~Spiropoulos (1993).
\newblock Prediction under the influence: {B}ayes linear influence diagrams for
  prediction in a large brewery.
\newblock {\em The Statistician\/}~{\em 42\/}(2), 445--459.

\bibitem[\protect\citeauthoryear{Goldstein and Wooff}{Goldstein and
  Wooff}{1995}]{gwblincomp}
Goldstein, M. and D.~Wooff (1995).
\newblock Bayes linear computation: concepts, implementation and programming
  environment.
\newblock {\em Statistics and Computing, to appear\/}.

\bibitem[\protect\citeauthoryear{Haff}{Haff}{1980}]{lhcovmat}
Haff, L. (1980).
\newblock Empirical bayes estimation of the multivariate normal covariance
  matrix.
\newblock {\em Ann. Statist.\/}~{\em 8}, 586--597.

\bibitem[\protect\citeauthoryear{Leonard and Hsu}{Leonard and
  Hsu}{1992}]{tlcovmat}
Leonard, T. and J.~Hsu (1992).
\newblock Bayesian inference for a covariance matrix.
\newblock {\em Ann. Statist.\/}~{\em 20}, 1669--1696.

\bibitem[\protect\citeauthoryear{Wooff}{Wooff}{1992}]{bdworks}
Wooff, D. (1992).
\newblock {[B/D]} works.
\newblock In J.-M. Bernardo et~al. (Eds.), {\em {B}ayesian Statistics 4}, pp.\
  851--859. Oxford University Press.

\end{thebibliography}

\end{document}